\documentclass{aa}

\usepackage{rotate,psfig}

\begin{document}

\thesaurus{06                   
               (08.09.2 SS433;                  
                09.10.1;                        
                13.18.5;                        
                       )} 

\title{The inner radio jet region and the complex environment of SS433}

\author{Z.~Paragi\inst{1,2}
\and
R.~C.~Vermeulen\inst{3}
\and
I. Fejes\inst{1}
\and
R.~T.~Schilizzi\inst{2,4}
\and
R.E.~Spencer\inst{5}
\and
A.~M.~Stirling\inst{5}
}

\offprints{Z. Paragi, 1st address (paragi@sgo.fomi.hu)}

\institute{F\"OMI Satellite Geodetic Observatory, P.O. Box 546, H-1373 Budapest, Hungary
\and
Joint Institute for VLBI in Europe, P.O. Box 2, 7990 AA
Dwingeloo, The Netherlands 
\and
Netherlands Foundation for Research in Astronomy, P.O. Box 2, 7990 AA
Dwingeloo, The Netherlands 
\and
Leiden Observatory, P.O. Box 9513, 2300 RA Leiden, The Netherlands
\and 
Nuffield Radio Astronomy Laboratory, Jodrell Bank, Macclesfield, Cheshire, SK11 9DL, UK. 
}

\date{Received 15 April 1999 / 
      Accepted 28 June 1999}

\authorrunning{Z. Paragi et al.}
\titlerunning{The inner radio jet region and the complex environment of SS433}
\maketitle

\vspace{0.5cm}

%
%
%
%
%
%
%

\begin{abstract}
  
  We present multi-frequency VLBA+VLA observations of \object{SS433}
  at 1.6, 5 and 15~GHz. These observations provide the highest angular
  resolution radio spectral index maps ever made for this object.
  Motion of the components of \object{SS433} during the observation is
  detected. In addition to the usual VLBI jet structure, we detect two
  radio components in the system at an anomalous position angle. These
  newly discovered radio emitting regions might be related to a
  wind-like equatorial outflow or to an extension of the accretion
  disk. We show that the radio core component is bifurcated with a
  clear gap between the eastern and western wings of emission.
  Modelfitting of the precessing jets and the moving knots of
  \object{SS433} shows that the kinematic centre -- i.e. the binary --
  is in the gap between the western and eastern radio core components.
  Spectral properties and observed core position shifts suggest that
  we see a combined effect of synchrotron self-absorption and external
  free-free absorption in the innermost AU-scale region of the source.
  The spatial distribution of the ionized matter is probably not
  spherically symmetric around the binary, but could be disk-like.


\end{abstract}

\section{Introduction}

SS433 is an eclipsing binary star which ejects antiparallel jets at a near
relativistic velocity of 0.26c (e.g. Vermeulen \cite{RCV95}). The kinematics 
is revealed by optical Doppler-shifted lines (e.g. Margon \& Anderson
\cite{M&A89}), 
as well as radio maps e.g. the corkscrew-shaped trails due to precession (Hjellming \& Johnston \cite{H&J81}). The jets show complicated
precessing and nodding motions, they are collimated to better than 4$\degr$.
The velocity of ejection is constant to within a few percent. Radio flux
monitoring programmes showed that \object{SS433} has active and quiescent
periods (e.g. Fiedler et al. \cite{FIE87}). Knots in the jets moving at 0.26c
can be seen on MERLIN scales (e.g. Spencer \cite{RES79}, \cite{RES84}). Radio
structure on a scale of 10 to 50~mas was discovered in 1979 (Schilizzi et al.
\cite{RTS79}). Since then, many VLBI observations have been made with a variety
of arrays and observing frequencies which revealed that moving components are
present in both the active and quiescent periods of SS433 (e.g. Romney et al.
\cite{ROM87}; Fejes et al. \cite{FEJ88}; Vermeulen et al. \cite{RCV93}). These
observations also showed that the central part of the radio source has a
core-wing morphology which is not centre-brightened. The presence of a
brightening zone -- where the moving components become brighter -- at
50~milliarcsecond from the centre was discovered by Vermeulen et al.
(\cite{RCV93}).

We present VLBI observations of \object{SS433} at 1.6, 5 and 15~GHz. These
observations provide us with the highest resolution spectral index maps ever
made for this object. In addition to the usual VLBI structure, our maps show
several new features in the system. In Sect.~2. we describe the
observations and the data reduction process. In Sect.~3. the VLBI 
images and some related observational results are presented. We discuss our
results in Sect.~4.

\section{Observations, calibration and data reduction}

The observations took place on 6 May 1995 with the VLBA and a single element 
of the VLA at 1.6, 5 and 15~GHz. At this time the binary orbital phase was 
0.28 (where $\phi$~=~0 corresponds to the eclipse of the accretion disk by 
the normal star) while the precession phase (following the definition of
Vermeulen \cite{RCV89}) of the radio jets was 0.5, when the approaching jet
lies closest to the line of sight. \object{SS433} was observed during 10 hours
using left circular polarization and 16~MHz bandwidth. The experiment included
some phase-referencing scans with the background radio sources \object{1910+052} (at all frequencies) and
\object{1916+062} (5~GHz only). Frequencies were cycled between 1.6, 5 and
15~GHz every 6.5 minutes except during the phase reference scans.

 \begin{figure*}
 \centerline{
 \psfig{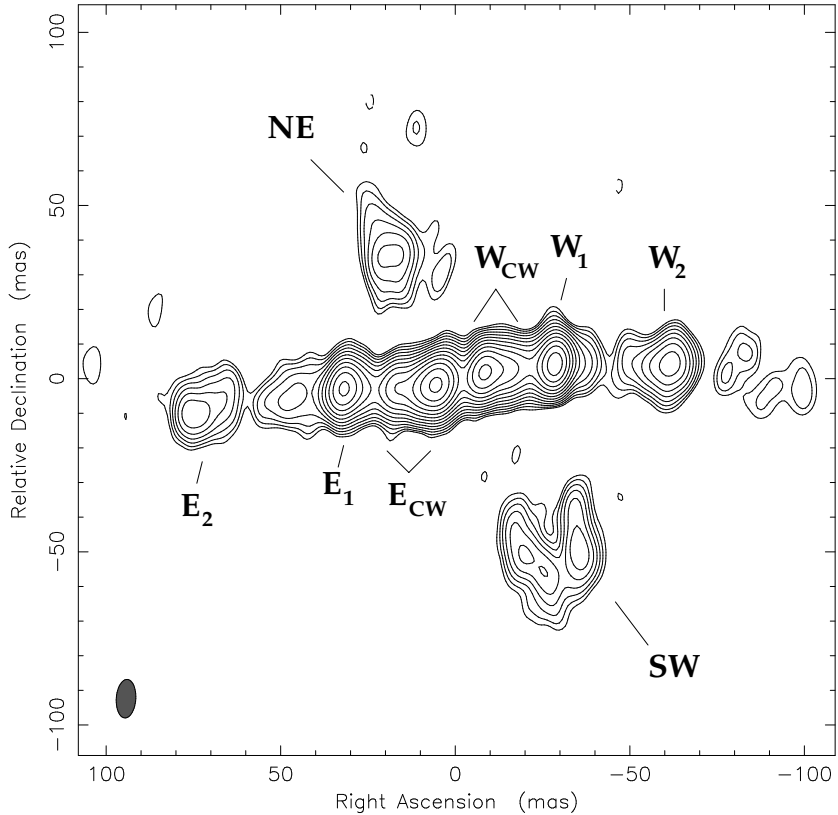}
 }
 \caption{Naturally weighted image of \object{SS433} at 1.6~GHz. For clarity we indicate some 
components mentioned in the text. Contour levels are $-$1.41, 1.41, 2.0, 2.83, 4.0, 
5.66, 8.0, 11.31, 16.0, 22.63, 32.0, 45.25, 64.0, 90.51~\% of the peak flux density of 
53.3~mJy/beam, the restoring beam is 11.14$\times$5.59~mas, PA=$-$3.8$\degr$}
 \label{18cm}
 \end{figure*}
 
 \begin{figure*}
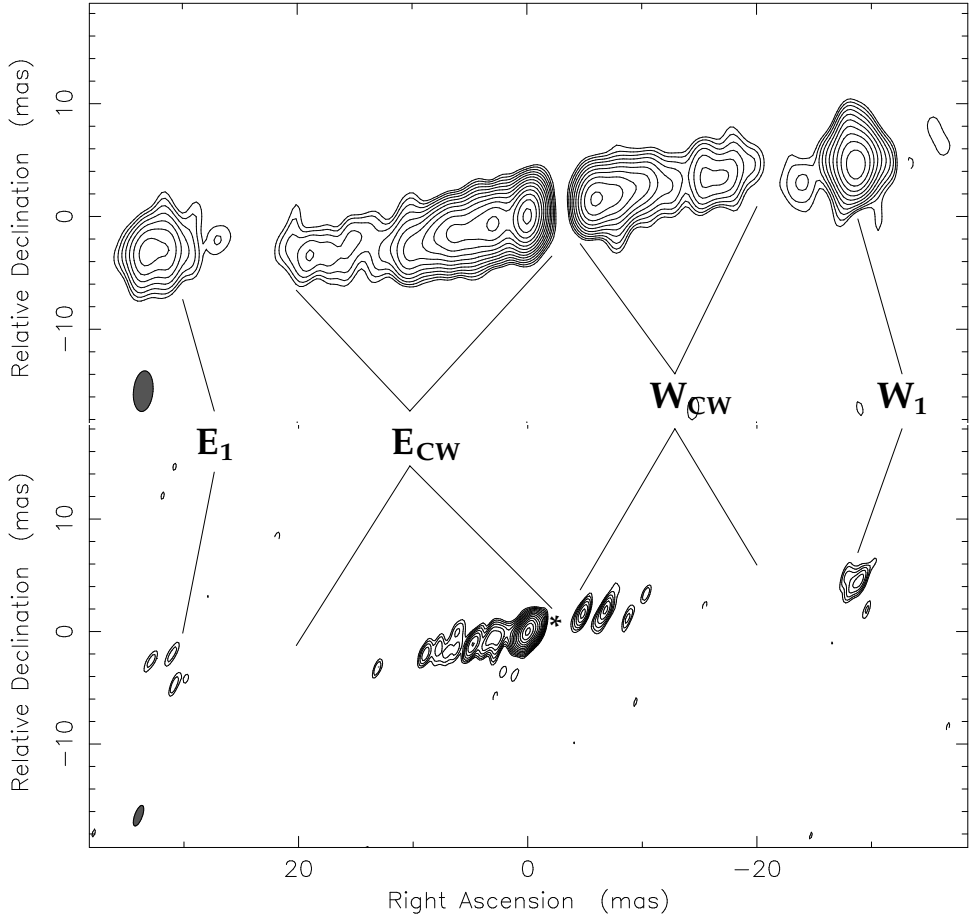

 \centerline{
 \psfig{file=h1498.f2a,width=13cm,bbllx=30pt,bblly=315pt,bburx=590pt,bbury=560pt,clip=}
 }
 \centerline{
 \psfig{file=h1498.f2b,width=13cm,bbllx=30pt,bblly=270pt,bburx=590pt,bbury=555pt,clip=}
 }
 \caption{Naturally weighted images of \object{SS433} at 5~GHz (upper image) and 15~GHz (lower image). 
For clarity we indicate some components mentioned in the text. The kinematic centre is indicated
by an asterisk in the 15~GHz image. Contour levels are $-$1.41, 1.41, 2.0, 2.83, 4.0, 5.66, 8.0,
11.31, 16.0, 22.63, 32.0, 45.25, 64.0, 90.51~\% of the peak flux densities of 36.2~mJy/beam (5~GHz) 
and 50.1~mJy/beam (15~GHz). The restoring beam is 3.62$\times$1.70~mas, PA=$-$5.6$\degr$ and
1.94$\times$0.67~mas, PA=$-$20.0$\degr$, respectively}
\label{6/2cm} 
 \end{figure*}

Initial data calibration was done using the NRAO AIPS package (Cotton
\cite{COT95}; Diamond \cite{DIA95}). We used measured antenna system
temperatures for amplitude calibration. There was a need to fringe-fit for
the delay and fringe rate for the outer antennas of the VLBA separately due 
to low signal-to-noise ratios on the longest baselines. Even with this 
method, solutions failed for the antennas MK and SC. We made a preliminary 
image using the rest of the array. After dividing the source model visibilities
into the dataset we repeated the fringe-fitting as described above, and were
able to obtain solutions for all antennas except SC at 15~GHz. The complex
bandpasses were calibrated using our fringe-finder sources \object{1803+784}
and \object{3C454.3}. After bandpass calibration, the data were averaged in
frequency in each IF and then in time (one minute integration time). Editing,
self-calibration and imaging was carried out using DIFMAP (Shepherd et al.
\cite{SHE94}). 

\section{Results}

\subsection{Images}

\begin{figure*}
 \centerline{
 \psfig{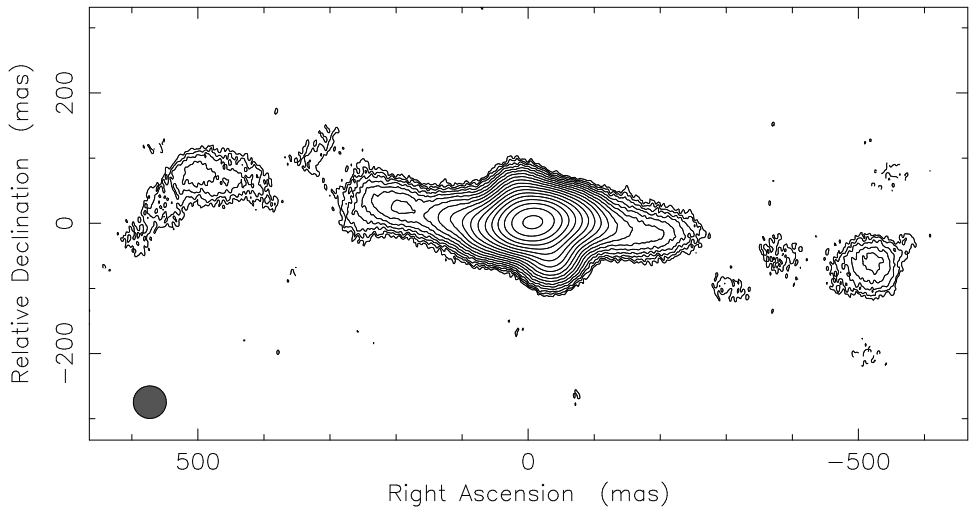}
 }
 \caption{The large scale structure of \object{SS433} at 1.6~GHz. Contour levels are $-$0.25, 0.25, 0.36, 0.50, 0.71, 1.0, 1.41, 2.0, 2.83, 4.0, 5.66, 8.0, 11.31, 16.0, 22.63, 32.0, 45.25, 64.0, 90.51~\% of the peak flux density of 248.3~mJy/beam. The restoring beam is 50~mas}
\label{18cm_lfv} 
\end{figure*}

We present the naturally weighted images of \object{SS433} at 1.6~GHz
(Fig.~\ref{18cm}.), 5~GHz, and 15~GHz (Fig.~\ref{6/2cm}.). At 1.6 and
5~GHz there is the well-known jet structure, oriented largely
East-West, which is roughly bi-symmetric in shape although not in
detailed flux density. We have labelled some of the more prominent
structures in accordance with our interpretation outlined below. By
using a 50~mas restoring beam, we were able to trace the jets out to
500~mas from the centre at 1.6~GHz (Fig.~\ref{18cm_lfv}.). The 1.6~GHz
image in addition contains extended features in locations oriented at
a large angle to the jets, where no emission has ever been seen
before. We have performed extensive tests to verify that these are not
artifacts of our data-reduction method. There is evidence for these
same regions at 5~GHz, but they are too faint and too extended to be
properly imaged from our data. The 15~GHz image is more asymmetric,
but shows the same general inner jet structure: E$_{1}$ and W$_{1}$
are identifiable with the same separation as at the lower frequencies,
and there is also emission at the location of the E$_{\mathrm cw}$ and
W$_{\mathrm cw}$ complexes. Clearly, all emission regions are
extended, and the details of their morphology are difficult to image
reliably at this high resolution.

Our resolution at 5 and 15~GHz was sufficient to detect the motion of some jet
components during the 12~hours observing time. Source structure changes did
not affect the final images significantly -- this was checked by splitting the
dataset into shorter timeranges. However, time-smearing may have affected the
dynamic range of our images, especially at 15~GHz.

\subsection{Alignment between frequencies}

Components E$_{1}$ and W$_{1}$, with their well-defined constant separation of
60~mas, are the most easily recognizable features between the three
frequencies, and we have used these for the relative alignment of the
images. We have been able to verify this relative alignment at 5 and
15~GHz by phase-referencing.  \object{SS433}  and \object{1910+052} (an
extragalactic background source located at an angular distance of 24$\arcmin$)
were observed alternately in short 1 min. (at 15~GHz) and 2 min. (at 5 and 1.6~GHz)
cycles during four phase-referencing scans. We fringe-fitted
\object{1910+052} using a point source model, and made an image. The
resulting model visibilities were divided into the dataset and we repeated
fringe-fitting. Delay, multiband-delay and fringe rate solutions -- corrected
for structural effects of the reference source -- were applied to SS433.
In this way images of \object{SS433} at different frequencies are overlaid 
naturally - assuming that the \object{1910+052} image centres at different
frequencies are exactly at the same place in the source. In agreement with our
first estimates, the 15 and 5~GHz images aligned well. Unfortunately, at 
1.6~GHz the reference source turned out not to have compact structure, perhaps
as a result of scattering, either in the supernova remnant \object{W50}, or
elsewhere along the line of sight through our Galaxy.

The ``absolute'' centre of ejection, where the binary stellar system
is located, is not obvious from the radio emission at any frequency,
and there is no centrally brightened ``core'' feature. For example, we
find that the brightest part of E$_{\mathrm cw}$ (on the side closest to the
centre), is 4~mas closer to E1 at 1.6 GHz than at 5 and 15~GHz.
Also, the separation between the inner parts of E$_{\mathrm cw}$ and W$_{\mathrm cw}$
decreases from 14~mas, to 6~mas, and then to 5~mas, in going from 1.6~GHz,
to 5~GHz, and then to 15~GHz.

However, we can use the well-established kinematic model of the
precessing jets of \object{SS433} (Margon \& Anderson \cite{M&A89}; 
Hjellming \& Johnston \cite{H&J81}; Vermeulen \cite{RCV89}) to find the centre. 
We believe that the features E$_{1}$-W$_{1}$ and E$_{2}$-W$_{2}$ are matched pairs, 
corresponding to successive events of ejection or knot formation which occurred 
simultaneously in both jets. Adopting the kinematic model parameters of 
Margon \& Anderson (\cite{M&A89}) we find ages of 3.5~days and 7~days for
E$_{1}$-W$_{1}$ and E$_{2}$-W$_{2}$, respectively.
The greater angular extent (i.e.\ apparent component spacing) on the
Eastern (approaching) side is in accordance with the light travel time
predictions of the kinematic model. Based on E$_{1}$ and W$_{1}$ in the 5~GHz
image, we derive that the most plausible location of the centre is
$\sim$2.5~mas W and $\sim 0.7$ mas N of the peak of E$_{\mathrm cw}$ at 5 GHz.

While our main interest was not to improve the accuracy of the
kinematic model parameters, we made a series of tests at all frequencies in 
which we varied the projected position angle of the precession cone axis 
in the plane of the sky, and independently the model precession phase; 
acceptable values of these parameters are well correlated  from our images. 
We find that the previously reported PA~=~100$\pm2\degr$ (e.g. Hjellming
\& Johnston \cite{H&J81}) fitted the structures well. Our main result is 
that the kinematic model centre is between E$_{\mathrm cw}$ and W$_{\mathrm cw}$, 
in the middle of the core-complex in the 15~GHz image (see Fig.~\ref{6/2cm}.).

\subsection{Spectral index maps}

\begin{figure*}
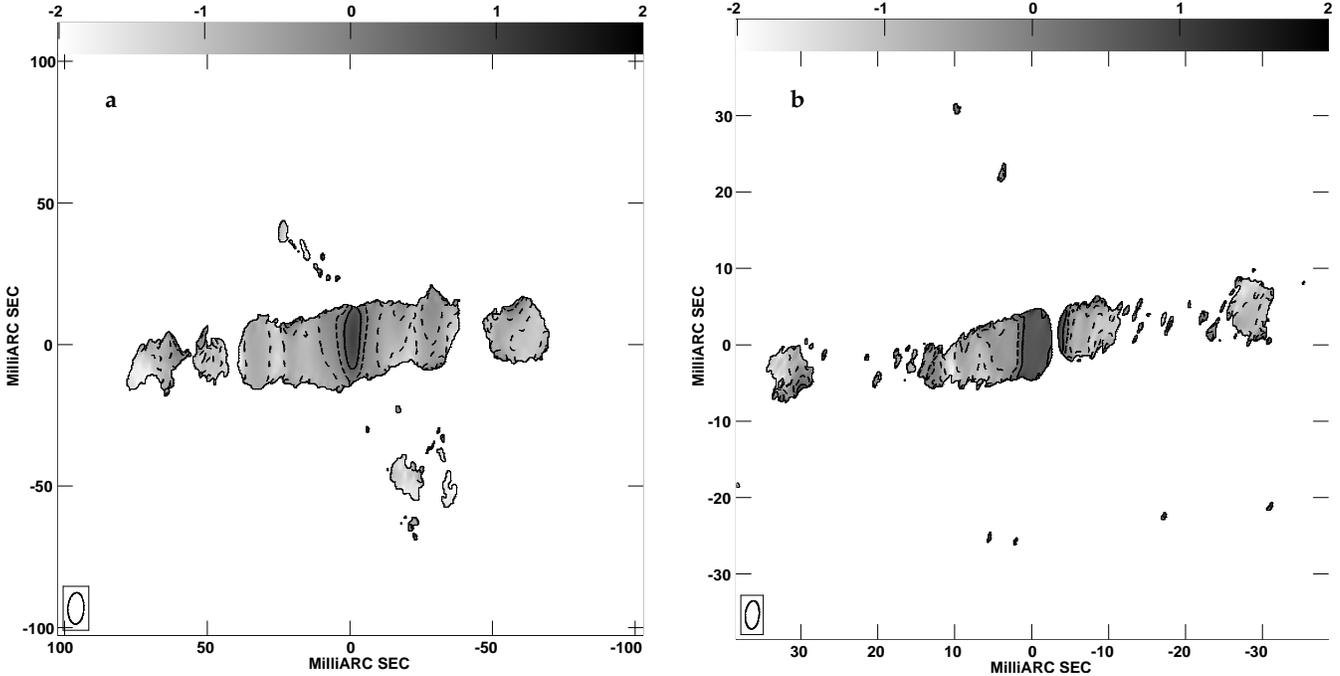

 \centerline{
 \psfig{file=h1498.f4a,width=9cm,bbllx=30pt,bblly=132pt,bburx=590pt,bbury=690pt,clip=}
 \psfig{file=h1498.f4b,width=9cm,bbllx=30pt,bblly=132pt,bburx=590pt,bbury=690pt,clip=}
 }
 \caption{Spectral index maps between {\bf a} 1.6 and 5~GHz {\bf b} 5 and
           15~GHz. There are hints of emission near SW and NE on the restored
           5~GHz image with spectral index $\sim-$1. E$_{1}$
           is also identified on the restored 15~GHz map. Contour levels in spectral index
           are $-$1, $-$0.5, 0, 0.5 ($S\propto\nu^{\alpha}$)}
 \label{spix}
 \end{figure*}

We used the relative image alignments as described in Sect.~3.2.
Shifting of the uv-data was performed by self-calibrating to model
files brought into alignment using the Caltech VLBI Package (Pearson\&
Readhead \cite{P&R84}) MODFIX program. Spectral index maps between
1.6-5~GHz and 5-15~GHz were made in AIPS after convolving the higher
frequency image with the lower frequency beam in both cases
(Fig.~\ref{spix}.).

The central region has an inverted spectrum, $\alpha_{1.6}^{5}=1$
($S\propto\nu^{\alpha}$ throughout this paper). Between 1.6 and 5~GHz, 
the jet spectral index steepens with distance from the core, decreasing 
below $-$1.0 at about 25~mas, while there are flatter regions, especially the
knots at 30~mas. These knots, E$_{1}$ and W$_{1}$, differ in spectral
properties, $\alpha_{1.6}^{5}=$~$-0.7$ for E$_{1}$, and $\alpha_{1.6}^{5}=-0.3$ for
W$_{1}$. In general, the two jets have similar spectral behaviour. This is not
true between 5 and 15~GHz, where the western jet steepens much faster, reaching
$\alpha_{5}^{15}=$~$-1$ less than 10~mas from the core. In the 5-15 GHz image we clearly resolve the core gap. The eastern part consists of a large, inverted area
($\alpha_{5}^{15}=$~$0.5-0.65$), while to the West only a thin inverted edge can
be seen ($\alpha_{5}^{15}$~$\sim$~0.5).

\subsection{Modelfitting and flux densities of the source components}

We performed modelfitting in DIFMAP. We divided the 5~GHz dataset into three 
parts in order to fit the positions of the source components during the
observation. The motion of E$_{1}$ and W$_{1}$ was detected, in agreement 
with the kinematic model. Our modelfitting results, however, are not accurate
enough to investigate the proper motion of these components in detail, because
the time interval is too short. We did not detect unambiguous motion in the
inner jets. Note that there are no well separated components in the inner 
part of the jets, which makes the modelfitting process difficult.

As the source structure cannot be described with a single set of
model components at all three frequencies, flux densities of the
components were determined from the images in AIPS. We defined eight
areas on the images (around E$_{\mathrm cw}$, W$_{\mathrm cw}$,
\ldots) in which flux densities were summed up. These are listed in
Table~\ref{modelfit}.

\begin{table*}
\caption{Flux densities of components. The coordinates of boxes in which the
flux density was integrated are also given}
\begin{tabular}{lrrrrrrr}
\hline
\noalign{\smallskip}

Component & \multicolumn{3}{c}{Flux density (mJy)} & \multicolumn{4}{c}{Box coordinates~(mas)}       \\
          &   1.6 GHz   &   5 GHz   &   15 GHz     &    $x_{1}$ &    $y_{1}$ &    $x_{2}$ &  $y_{2}$ \\
\noalign{\smallskip}
\hline
\noalign{\smallskip}
E$_{2}$   &   16.8      &    7.5    &     $-$      &       86.5 &   $-$23.5  &     58.5   &    18.0  \\
E$_{1}$   &   42.4      &   18.1    &     3.1      &       38.5 &   $-$18.0  &     23.0   &    12.5  \\
E$_{\mathrm cw}$  &  116.2      &  113.4    &   121.5      &       23.0 &   $-$16.0  &   $-$2.5   &    16.0  \\
W$_{\mathrm cw}$  &   76.1      &   44.3    &    17.4      &     $-$2.5 &   $-$13.0  &  $-$22.0   &    18.0  \\
W$_{1}$   &   56.6      &   33.0    &     9.5      &    $-$22.0 &   $-$ 9.5  &  $-$35.5   &    21.0  \\
W$_{2}$   &   25.6      &    9.9    &     $-$      &    $-$40.0 &   $-$11.0  &  $-$71.0   &    20.5  \\
NE        &   19.7      &    6.0    &     $-$      &       31.0 &      18.5  &   $-$1.0   &    58.5  \\
SW        &   32.6      &   10.5    &     $-$      &    $-$10.5 &   $-$72.5  &  $-$44.5   & $-$24.5  \\
\end{tabular}
\vspace{0.5cm}
\label{modelfit}
\end{table*}

\section{Discussion}

\subsection{The core region and the inner jets}

Our observations confirm the result of Vermeulen et al. (\cite{RCV93})
that the core area is not centre-brightened. For the first time, we
have the resolution to see a gap between the core-wings
(Fig.~\ref{18cm}. and \ref{6/2cm}.). Kinematic modelling -- based on
the jet curvature and symmetry of outer components -- showed that the
kinematic centre of the source is in the middle of the gap in the
15~GHz image.  The separation of the brightest parts of E$_{\mathrm
  cw}$ and W$_{\mathrm cw}$ increases with decreasing frequency as 
a result of synchrotron self-absorption, in rough accordance with
various different jet models, described below. However, the spectral
properties are quite dissimilar between E$_{\mathrm cw}$ and W$_{\mathrm cw}$.
It is obvious that their flux density ratio at any arbitrary frequency does 
not simply reflect the effect of differential Doppler-boosting as was thought to
be the case by Fejes (\cite{FEJ86}) and Vermeulen (\cite{RCV89}).
We argue below that the different spectra are caused by free-free absorption
in a medium which envelopes the various parts of the jets to different depths.

In the standard extragalactic jet model the distance of the self-absorbed radio core
from the central engine is proportional to $\nu^{-1}$ (Blandford \&
K\"onigl \cite{B&K79}); this leads to a predictable radio core position shift
between frequencies. In the case of continuous jets (steady emission pattern)
the arm-length ratio between the approaching and the receding jet is always
unity, and the central engine is simply located mid-way between the radio
"cores" in the two jets. Following the formalism of Lobanov (\cite{LOB98}), 
the measure of core position shift can be calculated by: 
\begin{equation}
\Omega_{r \nu} = 4.85 \cdot 10^{-9}\, \Delta r_{\mathrm{mas}} D \, \frac{\nu_{1}^{1/k_{\mathrm{r}}} \nu_{2}^{1/k_{\mathrm{r}}}}{\nu_{2}^{1/k_{\mathrm{r}}} - \nu_{1}^{1/k_{\mathrm{r}}}},
\label{pos-offset}
\end{equation}
where $\Delta r_{\mathrm{mas}}$ is the observed core position shift between observing frequencies
$\nu_{1}$ and $\nu_{2}$ (given in Hz), $D$ is the distance to the source in parsecs.
The parameter $k_{\mathrm{r}}$ is model dependent, and was assumed to be unity. 
The equipartition magnetic field in the jet at 1~pc from the central engine can be written (Lobanov \cite{LOB98}):
\begin{equation}
B_{1} = 2.92 \cdot 10^{-9}\, [ \frac{\Omega_{r \nu}^{3}}{ k_{\mathrm{e}} \delta_{\mathrm{j}}^{2}\, \phi \, \mathrm{sin}^{5}\theta} ]^{1/4}
\end{equation}
where $\delta_{\mathrm{j}}$ is the jet doppler factor, $\phi$ is the
jet opening angle and $\theta$ is the jet viewing angle. Using $B
\propto r^{-1}$ with $k_{\mathrm{e}}=1$, substituting $\phi=4\degr$
for the jet opening angle, and calculating the parameters dependent on
orientation ($\theta$ and $\delta_{\mathrm{j}})$ for precession phase
0.5, we find that the observed $\sim$4~mas core position shift in the
approaching jet side between 1.6 and 5~GHz results in an
equipartition magnetic field strength $B=0.4$ Gauss at 35~AU from
the central engine, where the 1.6~GHz core component is observed.

As the separation of E$_{\mathrm cw}$ and W$_{\mathrm cw}$ between 1.6
and 5 GHz is proportional to $\nu^{-1}$, and we obtained a reasonable
field strength in our calculations, we conclude that the core-wings
at 1.6~GHz are mainly self-absorbed. This seems to be in agreement
with the fact that the integrated flux density of E$_{\mathrm cw}$ is
nearly constant (Table~\ref{modelfit}.), i.e.  the overall spectral
index of the region is flat, as expected in case of synchrotron
self-absorption. The peak brightness of the approaching jet at 15~GHz,
however, does not seem to be high enough; the spectrum of a resolved,
self-absorbed component would be more inverted.  Moreover, the
position shift between the 5~GHz and 15~GHz core components seems to be
zero, or at least is smaller then our errors in aligning the images,
which is estimated to be a few tenth of mas. We interpret the absence
of core position shifts at high frequencies and the relative faintness
of the high frequency near-in core components in terms of additional 
free-free absorption.  Increasing column depth of thermal electrons may 
results in a situation where the innermost part of E$_{\mathrm cw}$ at 
15~GHz is absorbed so that the peak brightness is shifted outward from 
the central engine. Similarly, the brightest parts of the receding jet
(located consecutively closer to the central engine with increasing
frequency) are absorbed, and so the overall spectrum of W$_{\mathrm
  cw}$ remains steep from 1.6~GHz to 15~GHz.

We also compared the observed jet intensity profiles (i.e. the jet
brightness distribution as a function of distance to the central
engine) to the conical jet model of Hjellming \& Johnston
(\cite{H&J88}). In this model the jet is assumed to expand 
adiabatically and the expansion is dominated by lateral motions 
of the flow. We also assumed that the expansion is free, and not
slowed down by the interstellar medium.
In agreement with our interpretation above, we obtained
good first order agreement at 1.6~GHz between the shape of the observed and model
jet intensity profiles. This further supports that E$_{\mathrm cw}$
and W$_{\mathrm cw}$ are regions in transition from optically thick to
optically thin regimes.  The effect of Doppler-boosting and
de-boosting of the approaching and receding jet sides was also as
expected. However, this is not the case at 5~GHz and 15~GHz.  We
scaled the model to the 1.6 GHz data, then estimated the additional
free-free absorption required to explain the observed jet profiles at
higher frequencies. As this is very sensitive to the alignment of the
model with the data in position and the relative alignment of the jet
intensity profiles between the frequencies -- keeping in mind that we
have a few tenth of mas uncertainty in aligning the images --, the
values determined below must be considered as rough estimates. Note
also that these values are specific to the Hjellming \& Johnston
(\cite{H&J88}) adiabatic jet model.

 \begin{figure}
 \centerline{
 \psfig{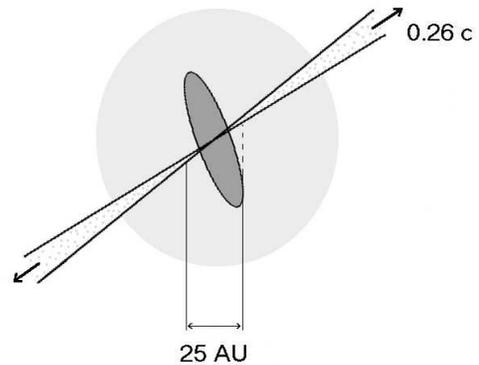}
 }
 \caption{A schematic view of the free-free absorbing medium around SS433. 
          The density of free electrons increases sharply towards the central
          engine. At 15~GHz the absorbing cloud becomes optically thick at a
          projected distance of $\sim$12.5~AU to the central engine (indicated by the inner shaded
          area), where $\tau_{\mathrm{ff}}$ exceeds 3 for the receding jet. The large
          asymmetry between the two sides and the sudden increase in optical 
          depths is most probably due to a disk-like geometry} 
 \label{discabs} 
 \end{figure}

The optical depths required to explain the observed peak flux
densities at 5~GHz are estimated to be $\tau_{\mathrm{ff}}=1$ and 2 at
the approaching and the receding side, respectively. As there is a
difference in the optical depths on the two sides, this additional
absorption cannot be intrinsic to the jets.  The required density 
of free electrons to produce $\tau_{\mathrm{ff}}=1$ at 5~GHz
is $n_{\mathrm e}=2.2\times10^{5}$ cm$^{-3}$  (assuming a thermal 
component with $T \sim10^{4}$~K, and a characteristic pathlength 
$L$=30~AU through the absorbing medium). The
relative faintness of E$_{\mathrm cw}$ at 15~GHz and the fact that
W$_{\mathrm cw}$ is absorbed almost completely suggest that the column
density increases toward the central engine.  The almost completely
absorbed peak of W$_{\mathrm cw}$ component at 15~GHz requires
$n_{\mathrm e}\sim1.2\times10^{6}$ cm$^{-3}$ or higher. We could not 
interpret our observations as the result of free-free absorption in 
a spherically symmetric stellar wind of the early-type star 
(Stirling et al.  \cite{AMS97}). Assuming a spherically symmetric 
configuration and e.g. $n_{\mathrm e} \propto r^{-3}$ dependence
of the free electron density on the distance to the binary, the 
differences in the integrated emission measures cannot explain the 
observed high difference in the optical depths between the two sides.
Instead we suggest the possibility of a disk-like geometry for the
absorbing medium (Fig.~\ref{discabs}). Note that this region is much
larger than the accretion disk. The projected size of the gap between
E$_{\mathrm cw}$ and W$_{\mathrm cw}$ is 25~AU, whereas the binary
system is generally thought to be $\sim$~1~AU in size. The spherically
symmetric versus equatorially enhanced stellar wind scenarios will be
studied in multi-epoch multi-frequency monitoring observations
spanning various phases of the precession cycle of \object{SS433}
which allow different lines of sight to be probed.

\subsection{The anomalous emission regions NE and SW}

Random projected velocity deviations of about 5000~km/s are frequently
observed in the jets of \object{SS433}; these have in the past been
successfully explained by deviations of only a few degrees in the
pointing angle of the jets ("jitter").  Detections of anomalous radio
emission components have been reported in some cases.  Modelfitting of
1981 single-baseline Effelsberg-Westerbork data (Romney et al.
\cite{ROM87}) indicated an elongated structure straddling the core in
PA~62\degr \, which is beyond the kinematic model cone. Spencer \&
Waggett (\cite{S&W84}) showed anomalous radio emission 15-20\degr \,
from the predicted locus in a 1982 series of EVN 5~GHz images. Jowett \&
Spencer (\cite{J&S95}) found a pair of knots apparently moving at only
half the predicted jet velocity in a 1991 and 1992 series of 5 GHz
MERLIN.

Our observation of \object{SS433} shows two faint, extended emission regions 
to the NE and SW from the central engine -- assumed to be located in between
E$_{\mathrm cw}$ and W$_{\mathrm cw}$ components as explained above -- extending 
in PA$\sim 30\pm 20$\degr\, over a distance of 30~mas to 70~mas in the 1.6~GHz image 
(Fig.~\ref{18cm}.). They are quite distinct from the "normal" jets, which are obviously 
also present. Features this far away in position angle from the jets have never 
been observed before. 

The implied brightness temperature of 
10$^{7}$~K suggests non-thermal radiation.
Should the NE and SW components be moving
approximately along the plane of the sky at 0.26c (like the jets), 
then these newly discovered features would have an age of a few days. 
But because this would imply an unprecedentedly large deviation from 
the kinematic model, because there were no noteworthy radio events in the source right 
before the observation (Fender et al. \cite {FEN97}),
and, most importantly, because we have seen the features again in more recent data, 
with similar disposition (Paragi et al. in preparation),
we believe that NE and SW are probably not anomalous knots which can be associated with the jets, 
ejected at near relativistic speeds. 
Instead, we believe these radio features are longer-lived, perhaps even permanent regions of emission extended along the equatorial plane of the binary system. We think they have not been previously detected because our VLBI array had an unprecedented sensitivity to extended, low surface brightness emission.

If we associate the anomalous components with a quasi-equatorial outflow from the system
with typical early-type stellar wind speeds (in the order of 1000~km/s), we may observe
the position angle of the flow to be correlated with the precessional phase. The possibility
of such an outflow or extended disk around \object{SS433} have been reported by several
authors as summarized below.

The asymmetric and variable shape of the optical lightcurve of \object{SS433} was explained
by Zwitter et al. (\cite{ZWI91}) as being due to the effect of an optically thick  disk-like outflow of matter extending more or less radially from the "slaved" accretion disk which is thought to surround the compact object in \object{SS433}. Slaved accretion disk models, in which different parts are in different planes because the matter is slaved to the rotation of the companion on which it originated, have already been invoked for \object{SS433} by several authors (e.g. van den Heuvel \cite{HEU81}) soon after its discovery.

Zwitter et al. (\cite{ZWI91}) proposed a relatively large ($\sim$20\degr) opening angle to the {\it excretion flow}, centred close to the orbital plane of the binary system. 
The presence of a very extended disk around SS433 (outer radius 1-3 arcmin, which translates to 
a few parsecs at the distance of the source) 
with an opening angle as large as $\sim$60\degr\ based on geometrical considerations
was invoked by Fabrika (\cite{FAB93}).

The properties of Doppler shifted X-ray lines as a function of the precession phase 
observed by Brinkmann et al. (\cite{BRI91}) and Kotani et al. (\cite{KOT96}) also indicate 
the need for a slaved disk model. In their model, there is an extended rim of the accretion disk.
However, the outer part of this disk must turn toward the orbital plane. So the opening angle of 
the extended envelope must be less than determined by Fabrika (\cite{FAB93}).

According to numerical simulations by Sawada et al. (\cite{SMH86}) 
a considerable fraction of the transferring gas is
lost in a binary system through the L$_{2}$ Lagrangian point,
behind the compact star. This results in an expanding envelope at PA=10\degr\ .
Numerical
simulations applied directly to SS433 also suggested the existence of an equatorial
outflow due to spiral shocks in the accretion disk (Chakrabarti \& Matsuda
\cite{C&M92}). This quasi-stationary spiral shock structure can also explain
the subday variabilities observed in the optical part of the spectrum
(Chakrabarti \& Matsuda \cite{C&M92}). 

We interpret NE and SW as the manifestation of the excretion disk in
the radio regime. During episodes of enhanced mass transfer from the
companion to the compact object, bright components emerge from the
core-complex (Vermeulen et al. \cite{RCV93}). This may also lead to
enhanced ejection into the excretion flow through the L$_{2}$
Lagrangian point, and develop shock waves into the ISM. Relativistic
electrons can be produced in the shock fronts that are responsible for
the observed non-thermal radiation.  The observed position angle and
the large extent of NE and SW are in agreement with the slaved disk
model discussed above. 

\section{Conclusion}

We have shown that \object{SS433} is active even in its ``low'' state.
There is a continuous inner jet region, and moving pairs of blobs are
present.  The eastern and western part of the core-complex is
separated by a gap, which is in fact the kinematic model centre. There
are fainter extended regions not connected directly to the moving jets
of the source. We overview the models that explain the various
activities observed in the system. We find that the slaved accretion
disk scenario - in general - is in agreement with our observations.
However, many questions remain to be answered. The asymmetry within
the core implies that the radio emission and absorption scenario is
not well established in SS433. On one hand, it is clear that there
must be an intrinsic asymmetry in the free-free absorbing medium. On
the other hand, we can not explain the spectral properties of the
core-wings with a simple model. Multifrequency monitoring of
\object{SS433} at different precessional phases will hopefully help us
to separate the effect of Doppler-boosting and viewing angle from
intrinsic properties, and to constrain the spatial extent of the
absorbing medium in the innermost part of the source. The appearance
of the extended disk also has to be monitored in further VLBI
observations at low frequencies.

\begin{acknowledgements}

ZP wishes to acknowledge support for this research by the European Union under 
contract CHGECT 920011, the Netherlands Organization for Scientific Research 
(NWO), the Hungarian Space Office, and hospitality of JIVE and NFRA where part of
this work has been carried out. We are grateful to the staff of the VLBA, the 
NRAO correlator for their support of our project. The National Radio Astronomy
Observatory is operated by Associated Universities, Inc. under a Cooperative
Agreement with the National Science Foundation. 

\end{acknowledgements}


\end{document}